\documentclass[journal]{IEEEtran}
\usepackage{amsmath}
\usepackage{amssymb}
\usepackage{color}
\usepackage{siunitx}
\usepackage{pstool,graphicx,tikz}

\def\bfh{{\mathbf h}}

\begin{document}
%

\title{Design and Performance Analysis of  
	Non-Coherent Detection Systems with Massive Receiver Arrays 
}

\author{\IEEEauthorblockN{
Lishuai Jing,
Elisabeth De Carvalho,
Petar Popovski,
{\`A}lex Oliveras Mart{\'i}nez \\
}
\IEEEauthorblockA{Section  Antennas, Propagation and Radio Networking (APNet) \\Department of Electronic Systems, 
Aalborg University, Denmark\\
email: {\{}lji, edc, petarp, aom{\}}@es.aau.dk}
}

\maketitle

\begin{abstract}
Harvesting the gain of a large number of antennas in a mmWave band has mainly been relying on the costly operation of channel state information (CSI) acquisition and cumbersome phase shifters. 
Recent works have started to investigate the possibility to use receivers based on energy detection (ED), where a single data stream is decoded based on the channel and noise energy. The asymptotic features of the massive receiver array lead to a system where the  impact of the noise becomes predictable due to a noise hardening effect. This in effect extends the communication range compared to the receiver with a small number of antennas, as the latter is limited by the unpredictability of the additive noise.
When the channel has a large number of spatial degrees of freedom, the system becomes robust to imperfect channel knowledge due to channel hardening. 
We propose two detection methods based on the instantaneous and average channel energy, respectively. Meanwhile, we design the detection thresholds based on the asymptotic properties of the received energy. 
Differently from existing works, we analyze the scaling law behavior of the symbol-error-rate (SER). When the instantaneous channel energy is known, the performance of ED approaches that of the coherent detection in high SNR scenarios. 
When the receiver relies on the average channel energy, our performance analysis is based on the exact SER, rather than an approximation. It is shown that the logarithm of SER decreases linearly as a function of the number of antennas. Additionally, a saturation appears at high SNR for PAM constellations of order larger than two, due to the uncertainty on the channel energy. Simulation results show that ED, with a much lower complexity, achieves promising performance both in Rayleigh fading channels and in sparse channels.

\end{abstract}



\IEEEpeerreviewmaketitle

\section{Introduction}

Deploying a large number of antennas at base stations or user devices can potentially introduce advantages over the traditional point-to-point MIMO systems, such as improved energy efficiency, interference suppression and more reliable transmission \cite{F.RusekD.PerssonKiongLauBuonE.G.LarssonT.L.MarzettaO.EdforsandF.Tufvesson2013} \cite{Lu2014}. One key requirement towards providing these benefits is to obtain an accurate channel state information (CSI) \cite{Lu2014}. However, acquiring CSI can be challenging due to channel aging or pilot contamination \cite{Lu2014,TruongHeath2013}. Without a reliable CSI, especially at low signal-to-noise ratio (SNR) regimes, coherent detection suffers from inferior decoding performance. 

Energy detection (ED) is a non-coherent method that offers a sub-optimal, but low complexity and power efficient solution compared to coherent detection \cite{Witrisal2009}. With ED, the transmitted symbols are decoded based on the envelope of the received signals. Therefore, exact value of the channel coefficients are not necessary. However, since the detection/decoding is performed based on the signal energy, the system should use non-negative
signal constellations. For example, non-negative pulse amplitude modulation (PAM) constellations have been documented for two different wireless standards for Millimeter-wave (mmWave) short-range communication, ECMA-$387$ and IEEE$802.15.3$c~\cite{Yong2010,Baykas2011}, respectively.

For single antenna systems, various aspects of PAM-ED systems have been extensively investigated in~\cite{Urkowitz1967a,Paquelet2004, Anttonen2009,Moorfeld2009,Wang2011a}. 
In \cite{Urkowitz1967a}, theoretical fundamentals of ED for communication systems are provided with an emphasis on the analysis of the receiver characteristics. 
Simple and efficient ED receivers are proposed and their performance is analyzed in impulse radio based systems~\cite{Paquelet2004} and multilevel PAM systems \cite{Anttonen2009,Moorfeld2009}. In order to design the optimal constellation, \cite{Moorfeld2009} takes the symbol error rate (SER) of each constellation as the objective function. Typically, the SER or receiver operating characteristics are limited by the presence of noise, since both the signal and noise energy are collected at the receiver. Thus, applications of ED with a small number of antennas are attractive to short-range communications in which high SNR can often be guaranteed. The multi-stage weighted ED receiver studied in~\cite{Wang2011a} aims at tackling the  issue of noise accumulation. The proposed method relaxes, to some extent, the requirement for high SNR; however, it introduces computational complexity on acquiring the desired parameters, e.g. optimal weights and channel energy.

Regarding systems with a large number of antennas, ED was proposed for mmWave communications in~\cite{Andersen2009}, which studies the range extension in the absence of additive noise.
Asymptotic treatments similar to the one described in~\cite{Andersen2009} have been carried out in  \cite{Chowdhury2014,A.ManolakosM.ChowdhuryandA.J.Goldsmith2014,Manolakos2015,Martinez2014} and \cite{Hammouda2015}.
In~\cite{Chowdhury2014,A.ManolakosM.ChowdhuryandA.J.Goldsmith2014,Manolakos2015}, non-coherent detection is proposed which does not require the instantaneous CSI, but rather the channel statistics. Utilizing an upper bound on the SER, the authors in \cite{Manolakos2015} 
optimize the input constellation assuming a fixed size constellation and 
different levels of uncertainty on the channel statistics.
The SER performance that results from the proposed constellation design is quite sensitive to the knowledge of the channel statistics. Using Gaussian approximations for the channel and noise energy, a constellation design is proposed using the average channel and noise energy in \cite{Hammouda2015}. In \cite{Jing2015c}, the authors propose information-theoretic bounds based on Gaussian approximations on the channel and noise energy. The proposed bounds are shown to be tight at both low and high SNR regions.


In this work, we describe a PAM-ED transceiver in which the transmission of a single data stream is processed by a large number of receive antennas, see Fig.~\ref{fig:Collector}. The PAM symbol $x$ is detected based on the addition of the signal energy $|y_i|^2$ from all the $M$ receive antennas, or equivalently, by calculating the arithmetic average $\frac{\sum_{i = 1}^{M}\vert y_i \vert^2}{M}$. In the case when the receiver is based on the average channel energy, one key difference from the two closest works \cite{Manolakos2015} and \cite{Hammouda2015} is that in this work we base our optimization on the \emph{exact} SER, rather than approximations as in \cite{Manolakos2015} and \cite{Hammouda2015}. In that sense, our work provide a baseline to which any work  based on approximations should be compared. The other important contribution of this paper is that we devise two different methods that make use of the law of large numbers and the central limit theorem resulting from the presence of a large number of antennas $M$. 
The \emph{first} method, denoted as I-ED, is based on the instantaneous channel energy and relies on noise hardening. The central limit theorem allows a Gaussian approximation of the instantaneous received signal energy, which results in a closed-form expression of the detection thresholds. Moreover, this approximation allows for a performance characterization and analysis at high SNR regimes. We prove that, for a positive PAM, the performance of I-ED asymptotically approaches that of coherent detection. 
The \emph{second} method, denoted as A-ED, is based on the average channel energy and relies on both channel and noise hardening. A-ED employs solely the second-order statistics of the channel and hence is robust to channel knowledge uncertainty or small-scale movements of the users. However, in some practical scenarios, where the number of degrees of freedom (DoFs) in the channel is limited, the performance of A-ED becomes poor,  whereas I-ED maintains promising performance. 

When A-ED is utilized, assuming that a very large number of DoF in the channel is available, we obtain expressions of the SER based on Chi-square cumulative distribution functions (cdfs). Note that these expressions do not rely on the Gaussian approximation and therefore should be treated as exact solutions, not approximations. This SER is further used to optimize the input signal constellation. 
Based on simulations, we compare the performance of I-ED and A-ED in sparse channels with limited number of DoFs. 
While A-ED exhibits poor performance, I-ED remains robust in such propagation environment and should be the method of choice for this case. 

In the following, we use  $(.)^H$, $E(.)$, and $||\cdot||_2$ to denote  Hermitian, expectation and the $L_2$ norm operators.

\section{System and Signal Model}
\label{sec:sysmod}

\begin{figure}[!h]
	\centering
	\includegraphics[width=0.98\columnwidth]{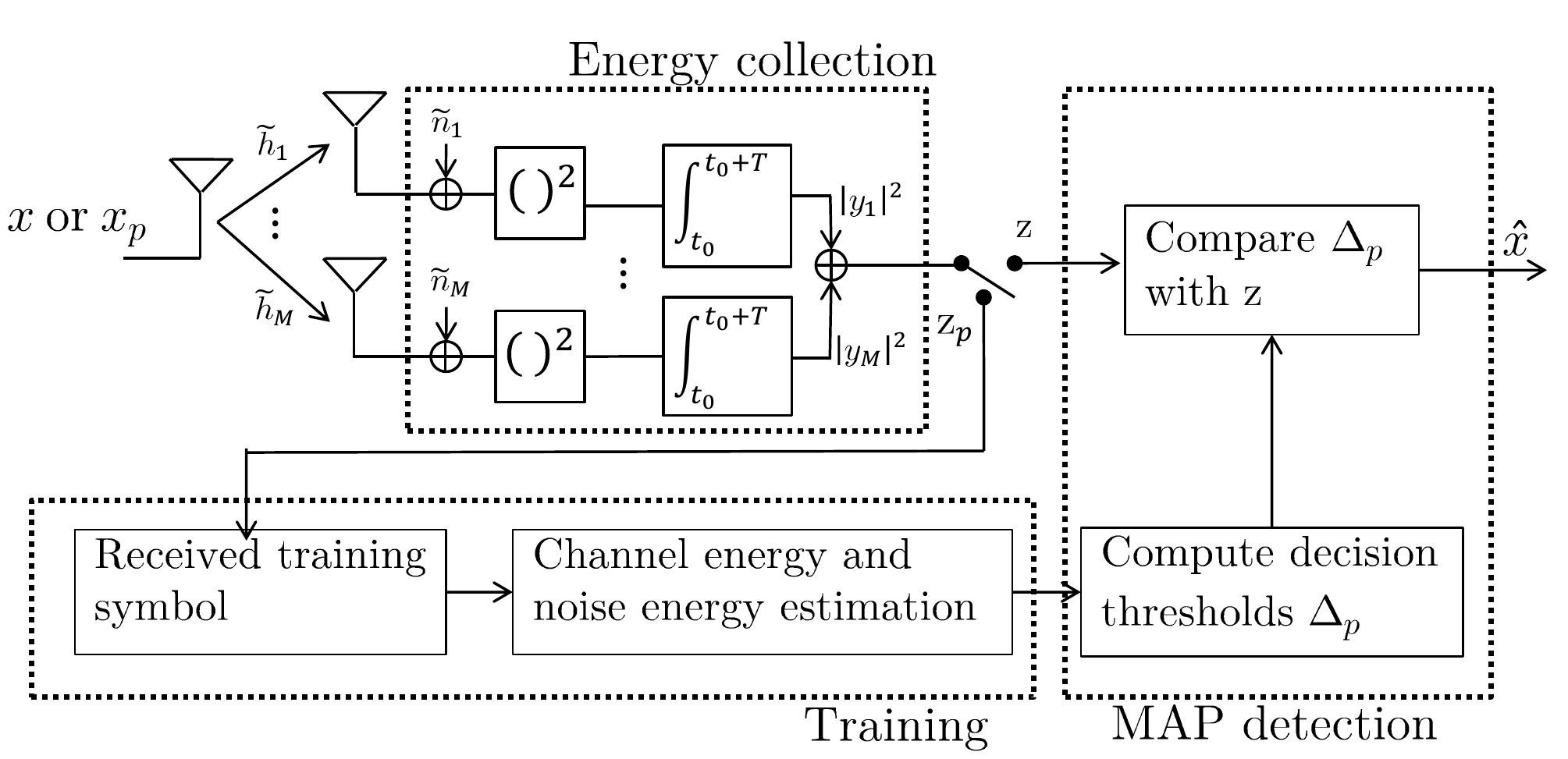}
	\caption{A simplified diagram for ED. Training symbols $x_p$ can be used to estimate the noise variance and channel energy. In data transmission mode, these estimates are used to compute thresholds based on the MAP/ML decision rule to decode data symbol $x$. Note that we use $h_i$ for $i = 1, \ldots, M$, in \eqref{yi} to denote the equivalent channel coefficients of $\tilde{h}_i$ in the analog domain. The same applies for the additive noise.}
	\label{fig:Collector}
\end{figure}

The considered transceiver structure is shown in Fig.~\ref{fig:Collector}. 
As one of the potential applications is mmWave communications, the transmitter and receiver are equipped with a large number of antennas. 
A single data stream is assumed to be transmitted and,  
as a  simplification, transmission in the uplink is performed from a single transmit antenna, see Fig.~\ref{fig:Collector}. Alternatively, the transmission can use multiple antennas but using a large beam (e.g. based on long term characteristics of the channel). 
Indeed, as the single data stream is non-coherently detected at the receiver, it is natural to assume that the phases of the channel coefficients are not available at the transmitter, such that  beamforming based on instantaneous CSI is not possible.

At the receiver's front end, the signal from each of the available $M$ receive antennas is filtered, squared, and integrated. The outputs from each antenna are summed up,  which is referred to as \emph{energy collection}; more details can be found in \cite{Urkowitz1967a}. 
{With an integration pefromed over a given time window $T$, the corresponding DoFs 
appear in the communication channel~\cite{Urkowitz1967a}. Without loss of generality and for the ease of presentation,
we ignore those DoFs. Although they are not negligible, we
are primarily interested in the DoFs brought by the multiple
antennas. Thus, in the simplified scheme, the integrator is replaced by a sampler which takes a single sample of the received signal over the time window $T$. }

Assuming the guard interval is large enough such that inter-symbol interference is negligible~\cite{Anttonen2009}, the received signal at a certain time instant after the energy collector reads \cite{Martinez2014,Manolakos2015}
\begin{equation}\label{eq:main}
z = \frac{1}{M} \sum^{M}_{i = 1} \left| y_i \right|^2. 
\end{equation}
The division by $M$ is introduced for normalization and the received signal at the $i$th antenna is given by   
\begin{equation}\label{yi}
y_i = h_i x + n_i,
\end{equation}
where $h_i$ denotes the channel coefficient between the transmitter and the $i$th receiver antenna in the digital domain and $x$ is the unknown transmit symbol. We assume that $h_i$, for $i = 1, \ldots, M$, has zero mean and variance $\sigma^2_{h}$. The additive noise contribution, $n_i$, for $i = 1, \ldots, M$,  is circular complex Gaussian with zero mean and variance $\sigma^{2}_n$.  
Since the signal energy needs to be collected at the receiver, the transmitted symbol $x$ is selected from the non-negative constellation set
\begin{equation}\label{Constellation}
x \in \{0, \sqrt{\epsilon_1}, \sqrt{\epsilon_2}, \dots, \sqrt{\epsilon_{P-1}}\},
\end{equation}
where $E[x^2] = 1$. The task is to decode the unknown symbol $x$ based on the observation $z$ resulting from  energy collection. 


\section{I-ED and A-ED:  Asymptotic Properties  and Channel Assumptions}

\label{sec:EDLA}

The two ED methods described in this paper are called \emph{Instantaneous channel energy based ED (I-ED) }
and 
\emph{Average channel energy based ED (A-ED)}. 
In this section we examine  the asymptotic properties that are exploited by I-ED and A-ED  as well as the related channel assumptions. 
For this purpose, we reformulate \eqref{eq:main} as
\begin{equation} \label{EDoutput}
z =  \underbrace{\frac{1}{M} \| \mathbf{h}\|^2_2  }_{\text{\normalsize{$\varsigma_h$}}} x^2 +
\underbrace{\frac{1}{M} \| \mathbf{n}\|^2_2 }_{\text{\normalsize{$\varsigma_n$}}}
\; + \; 
\underbrace{
2 \frac{1}{M} \Re (\mathbf{h}^H \mathbf{n})}_{\text{\normalsize{$w$}}} x. 
\end{equation}

 \subsection{Instantaneous channel energy based ED (I-ED)}
 
 I-ED assumes that the instantaneous channel energy $\varsigma_h$ is perfectly known at the receiver. 
 I-ED  essentially exploits 
 the assumption that the noise at the receiver is independent and identically distributed (i.i.d.) across antennas, resulting in {\it noise hardening}. 
This assumption is well verified in general if  $\mathbf{n}$ is thermal noise, which is the assumption adopted in this paper. 
If, instead, $\mathbf{n}$ represents interference coming from a single user, then the DoFs available in the interference channel may be limited by the propagation properties. However, when multiple users are present as interfering signals, it is unlikely that they will occupy the same angular space.
Hence,  the more users are interfering, the more DoFs become available in general, enabling noise hardening.

Under the adopted assumptions, as $M \rightarrow + \infty$, $\varsigma_n$ converges to the noise variance $\sigma_n^2$, according to law of large numbers. Likewise, the third term $w$  becomes arbitrarily close to zero.  
Due to the noise hardening effect, in principle, the noise contribution can be 
removed from $z$ in (\ref{EDoutput}) provided that the noise term $\sigma^2_n$ is reliably estimated. 
ED methods with a small number of receive antennas suffer from the accumulation of noise in the energy collection process which restricts the communication reach. I-ED alleviates this problem and widens the application of ED beyond short-range communications to possibly cellular systems.  

Furthermore, according to the central limit theorem, the received signal $z$ in (\ref{EDoutput}) can be approximated as a non-centered Gaussian random variable (see Section~\ref{Approach1}). This approximation leads to convenient analysis. It is a notable advantage over ED with small number of antennas where  (\ref{EDoutput}) is modeled as a non-central Chi-square probability density function (pdf) for which the determination  of the detection thresholds 
requires numerical methods.

\subsection{Average channel energy based ED (A-ED)}
 
A-ED also exploits noise hardening and the convergence of $w$  to zero, hence inheriting the properties of the aforementioned I-ED. 
Furthermore,  the channel is assumed to have a Rayleigh distribution, such that $\mathbf{h} = [h_1, \ldots, h_M]^{T} \sim  {\cal C}{\cal N}(0, \sigma_h^2 I)$. Under this assumption, as $M \rightarrow + \infty$,
the instantaneous channel energy $\varsigma_h$ tends asymptotically to the average channel energy $\sigma_h^2$, referring to as \emph{channel hardening}. 
Note that  channel hardening  also holds for certain correlated channel models \cite{Couillet2011} or 
when the transmitter and receiver communicate in line-of-sight (LOS) conditions.
In A-ED, detection is performed based solely on the average channel energy  $\sigma_h^2$ and noise energy $\sigma_n^2$.

\vspace{1mm}

The limitation of A-ED resides in the channel hardening properties which puts strong requirements on the propagation conditions. 
For sparse channels, which contain a small number of propagation paths, or  cluster-based models with a small angular spread of the scattering clusters, channel hardening is only partial. In such a case, I-ED becomes the method of choice. Provided channel hardening can be guaranteed,  A-ED finds its prime interest for fast fading channels when the instantaneous channel energy cannot be tracked at the receiver. More details on channel energy estimation can be found in~\cite{Jing2016a}.

In the sequel, we assume that the noise energy $\sigma^{2}_n$ stays constant and can be perfectly estimated over a long observation window, e.g. when the system is in the idle mode.

\section{Instantaneous channel energy based ED (I-ED)}
\label{sec:ML}

In this section, we assume perfect knowledge of $\varsigma_h$ and $\sigma_n^2$ at the receiver and apply the maximum a posteriori (MAP) principle to decode the unknown symbol $x$. 
We set up the following hypothesis
$${\cal H}_p: \;\; x = \sqrt{\epsilon_p}, \;\; \text{for} \; p = 0, \ldots, P -1$$
with prior distribution $p({\cal H}_p) = p(x = \sqrt{\epsilon_p})$. Accordingly, the MAP decision rule is written as:
\begin{equation}\label{eq:MLD}
\frac{f(z | \varsigma_h , {\cal H}_p)}{f(z |\varsigma_h, {\cal H}_{p'})} \underset{{\cal H}_{p'}}{\overset{{\cal H}_p}{\gtrless}} \frac{p({\cal H}_{p'})}{p({\cal H}_p)} , \quad \forall \; p' \neq p,
\end{equation}
where $f(z | \varsigma_h , {\cal H}_p)$ is a non-central Chi-square  pdf of $z$ conditioned on  
$\varsigma_h$ and ${\cal H}_p$ in our problem. 

The decision threshold between two neighboring constellation points is denoted 
as $\Delta_{p}(\varsigma_h)$, for $p=0,\dots P-2$. Therefore, we have 
\begin{equation}
x = \begin{cases}
\epsilon_0  &  z < \Delta_{0}(\varsigma_h) \\
\epsilon_p  & \Delta_{p-1}(\varsigma_h) \leq z < \Delta_{p}(\varsigma_h) \text{ for } 1\leq p\leq P-2 \\
\epsilon_{P-1} & z \geq \Delta_{P-2}(\varsigma_h)
\end{cases}.
\end{equation}

Employing the decision rule, the probability of detecting $\epsilon_{p+1}$ while $\epsilon_{p}$ is transmitted  is 
denoted as $P_e^u(\varsigma_h, \epsilon_p)$ where 
the superscript ``$u$'' refers to the upper tail of $ f(z| \varsigma_h,\epsilon_p)$. 
In fact, $P_e^u(\varsigma_h, \epsilon_p)$ is a Marcum Q-function which is expressed as
\begin{align}\label{PeUInst}
P_e^u(\varsigma_h, \epsilon_p)  =  
 \int_{\Delta_p(\varsigma_h)}^{\infty}   f(z| \varsigma_h,\epsilon_p)  dz.
\end{align}
Similarly, the probability of detecting $\epsilon_{p-1}$ while $\epsilon_{p}$ is transmitted is 
denoted as $P_e^l(\varsigma_h, \epsilon_p)$ where 
the superscript ``$l$'' refers to the lower tail of $ f(z| \varsigma_h,\epsilon_p)$. 

Using (\ref{eq:MLD}) and the involved Chi-square pdfs, it is possible to find the detection thresholds via numerical methods. Furthermore, the SER averaged over the channel can be determined using (\ref{PeUInst}) and Monte-Carlo simulations. 
Next, we approximate $ f(z| \varsigma_h,\epsilon_p)$ by a non-centered Gaussian pdf to compute the detection thresholds and characterize the SER performance at high SNRs.

\subsection{Gaussian Approximation of $z$ Conditioned on $\varsigma_h$} \label{Approach1}

\begin{figure}[!h]
	\centering
	\includegraphics[width=0.95\columnwidth]{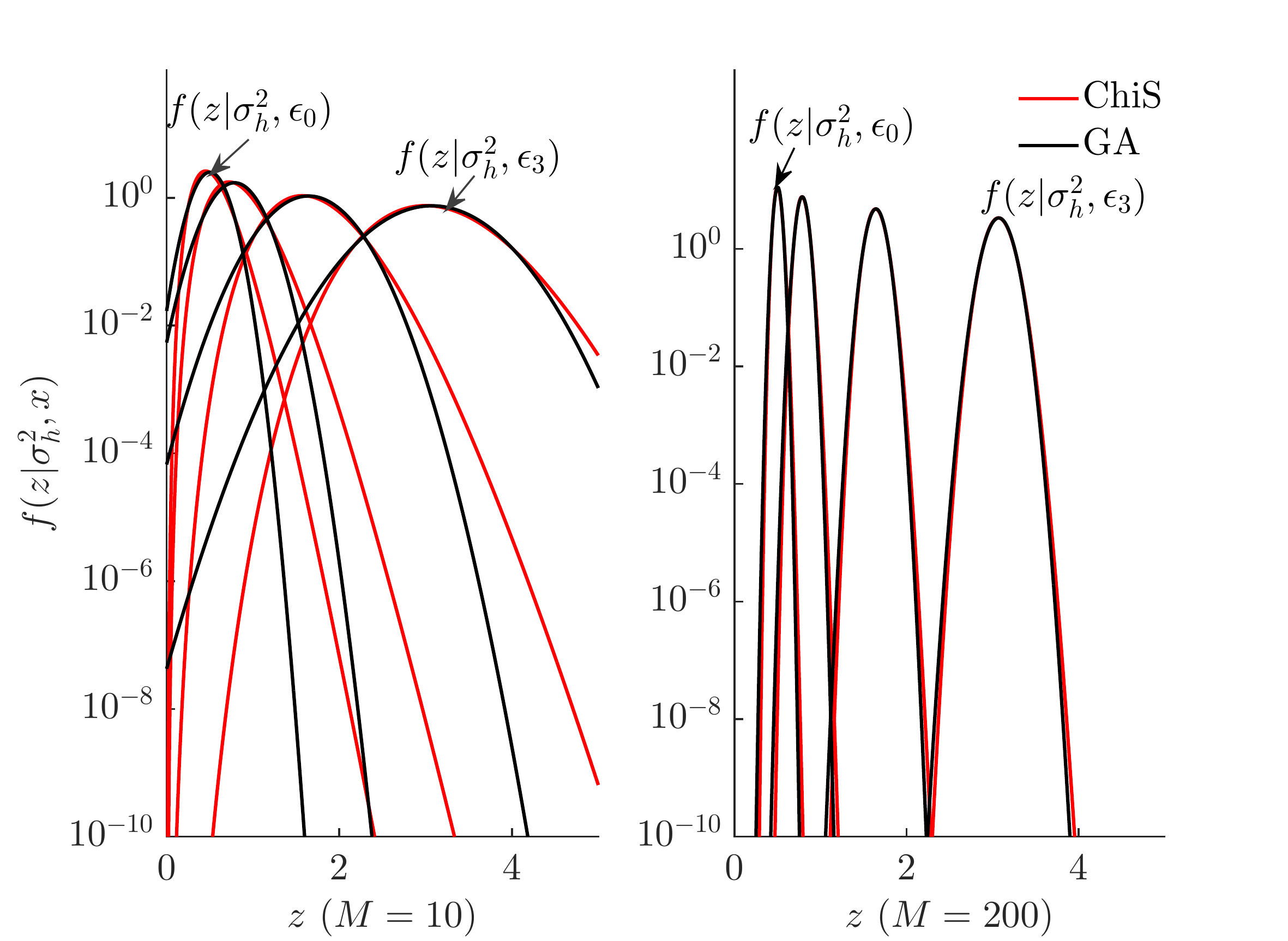}
	\caption{Conditional pdfs of $z$ in (\ref{eq:main}) expressed in logarithmic scale for a given channel realization with different number of receive antennas:  $ \frac{\sigma^2_h}{\sigma^2_n}= \SI{3}{\decibel}$ and $P = 4$. The legend ``ChiS'' symbols the Chi-square pdf $f(z|\varsigma_h=\sigma^2_h,x = \epsilon_p)$. We use ``GA'' to indicate the Gaussian pdf with the same first and second moments.}
	\label{fig:pdf200}
\end{figure}

Relying on the central limit theorem, $z$ in (\ref{EDoutput}) can be approximated 
as a non-centered Gaussian random variable when $M$ is large, i.e. 
$f(z|\varsigma_h,\epsilon_p) \sim  {\cal C}{\cal N}(\mu_z, \sigma^2_z)$ with the same first and second moments as the non-central Chi-square distributed random variable $z$ \cite{Martinez2014}:
\begin{equation}\label{zMean}
\mu_z  \! \left(\varsigma_h,\epsilon_p \right)= {\mu}_{z,p} = 
\varsigma_h \epsilon_p + \sigma_n^2,
\end{equation}
\begin{equation}\label{zVar}
{\sigma}^2_z \! \left(\varsigma_h,\epsilon_p \right) =  {\sigma}^2_{z,p} =  \frac{ 
  \sigma_n^2}{M}\left (  2 \varsigma_h \epsilon_p + \sigma_n^2 \right).
\end{equation}
The conditional pdfs $f(z|\varsigma_h, \epsilon_p)$ under a Rayleigh fading channel assumption are depicted in Fig. \ref{fig:pdf200} using a logarithmic scale. When $M$ is small, there is a noticeable difference between the exact pdf and its Gaussian approximation. However, when $M$ is large, the two pdfs become alike. Because the tails of the pdf are important for determining the SER performance, a good fit on the tails is observable.
In addition, the conditional variance \eqref{zVar} becomes larger as the energy level $\epsilon_p$ increases, which can be observed in the figure.


\subsection{Detection Thresholds} 

The decision thresholds from \eqref{eq:MLD} are computed based on the MAP rule using the Gaussian pdfs with moments expressed in \eqref{zMean} and \eqref{zVar}. 
With simple manipulations, $\Delta_p(\varsigma_h)$ is shown to be 
the positive root of the second-order polynomial equation 
\begin{equation}\label{eq:roots}
\boldsymbol{\vartheta}^{T} \boldsymbol{x} = 0, 
\end{equation}
where $\boldsymbol{x} = [x^2  \; \;  x \;\;   1]^{T}$ and 
$$\boldsymbol{\vartheta} = \begin{bmatrix}
{1}/{{\sigma}^2_{z,p}} - {1}/{{\sigma}^2_{z,p+1}} \\
2 \left( {\mu_{z,p}}/{{\sigma}^2_{z,p}}  - {\mu_{z,p+1}}/{{\sigma}^2_{z,p+1}}  \right) \\ 
\Big( {\mu_{z,p}^2}/{{\sigma}^2_{z,p}}  -  {\mu_{z,p+1}^2}/{{\sigma}^2_{z,p+1}} +\\ \log \left(
{{\sigma}^2_{z,p+1}} / {{\sigma}^2_{z,p}} \right)  + 2\log(\frac{p(\epsilon_{p+1})}{p( \epsilon_{p})}) \Big)
\end{bmatrix}. $$
We remark that the result reported in \cite{Martinez2014} is a special case of \eqref{eq:roots} by setting $p( \epsilon_{p}) = \frac{1}{P}$, for $p = 0, \ldots, P-1$.

\subsection{Performance Analysis}\label{Sec:BF}
 
We now assume that the detection thresholds are computed via \eqref{eq:roots} and characterize the SER performance.

\subsubsection{Performance Based on Gaussian Approximations}
Invoking the Gaussian approximations, 
$P_e^u(\varsigma_h, \epsilon_p)$ is approximated as: 
\begin{eqnarray}
P_e^u(\varsigma_h, \epsilon_p) 
 \approx  Q\left(\sqrt{\gamma_{u,p}} \right) \!, \;\;
\label{PePerChUPInst}
\end{eqnarray}
\normalsize
where the $Q$ function is defined as
$$Q(y) = \frac{1}{\sqrt{2 \pi}}\int_{y}^{+\infty} e^{-\frac{a^2}{2}} \mathrm{d} a$$ and $$\gamma_{u,p} = \left(
{\frac{\Delta_p(\varsigma_h) - \mu_z  \! 
		\left(\varsigma_h,\epsilon_p \right)}{\sigma_z  \! \left(\varsigma_h,\epsilon_p \right)}}\right)^2.$$
Similarly, $P_e^l(\varsigma_h, \epsilon_{p+1})$ is approximated as
\begin{eqnarray}\label{PePerChLowTailInst}
P_e^l(\varsigma_h, \epsilon_{p+1}) 
 \approx  Q\left( - \sqrt{\gamma_{l,p}} \right) \! \;\;
\end{eqnarray}
and $$\gamma_{l,p} = \left(
{\frac{\Delta_p(\varsigma_h) - \mu_z  \! \left(\varsigma_h,\epsilon_{p+1} \right)      }{\sigma_z  \! \left(\varsigma_h,\epsilon_{p+1} \right)}}\right)^2.$$
Note that 
$\gamma_{u,p}$ and $\gamma_{l,p}$ are important quantities as they can be interpreted as the post-processing SNR of the communication system. For systems with a small number of antennas, it is not possible to define such an SNR.  Accordingly, the SER reads 
\begin{equation}\label{PeInst}
P_e(\varsigma_h) \approx \sum_{p = 0}^{P-2} \left(Q(\sqrt{\gamma_{u,p}}) p(\epsilon_{p}) + Q(-\sqrt{\gamma_{l,p}})p(\epsilon_{p+1}) \right).
\end{equation}

\subsubsection{Scaling Behavior of SER at High SNR Regimes}\label{BF:analysis}
We show that I-ED asymptotically achieves the same performance as coherent detection (with a positive PAM constellation) and hence has a diversity order of $M$.

We define the instantaneous SNR as $\rho_h =  \varsigma_h/ {\sigma_n^2}$. 
When $\rho_h$ is large, based on \eqref{eq:roots},  we can show that 
\begin{equation}\label{PAMdeltaInst}
\begin{aligned}
\Delta_0(\varsigma_h) &=   
\sigma_n^2 \sqrt{\frac{\epsilon_1 \rho_h }{2}}   \\
\Delta_p(\varsigma_h) & =  \varsigma_h \sqrt{\epsilon_p} \sqrt{\epsilon_{p+1}}, \;\;  \text{for} \;p\neq 0.
\end{aligned}
\end{equation}	
Furthermore, the asymptotic moments are 
\begin{equation}\label{PAMMomentsInst}
\begin{aligned}
\mu_z  \! \left(\varsigma_h,\epsilon_0 \right) &=   
\sigma_n^2; \;\; \sigma^2_z  \! \left(\varsigma_h,\epsilon_0 \right) = \frac{\sigma_n^4}{M},  \\
\mu_z  \! \left(\varsigma_h,\epsilon_p \right) &=   
\varsigma_h \epsilon_p;  \;\; \sigma^2_z  \! \left(\varsigma_h,\epsilon_p \right) = 2\frac{\sigma_n^2 \varsigma_h \epsilon_p}{M} \;\;  \text{for} \;p\neq 0.
\end{aligned}
\end{equation}
Employing \eqref{PAMdeltaInst} and \eqref{PAMMomentsInst}, we prove that 
$\gamma_{l,p}$ and $ \gamma_{u,p}$ are equal and can be  expressed as:
\begin{equation}\label{GammaRelations}
\begin{aligned}
\gamma_{l,p} & = \gamma_{u,p} \approxeq M \rho_h {\kappa_p}, \;\;\;
\kappa_p = \left(\frac{\sqrt{\epsilon_{p+1}} - \sqrt{\epsilon_{p}}}{\sqrt{2}}\right)^2.
\end{aligned}	
\end{equation}
As a result, the probability of error $P_e^u(\varsigma_h, \epsilon_p)$ (and likewise $P_e^l(\varsigma_h, \epsilon_{p+1})$) reads
\begin{eqnarray}
P_e^u(\varsigma_h, \epsilon_p) 
 \approx  Q\left( \sqrt{\sum_{i=1}^M |h_i|^2  \frac{{\kappa_p}}{\sigma_n^2}}\right) 
\label{PePerChUPInstHighSNR}
\end{eqnarray}
The quantity 
$\gamma_{l,p} = \sum_{i=1}^M |h_i|^2  \frac{{\kappa_p}}{\sigma_n^2}$ can be interpreted as 
the post-processing SNR of a coherent (by matched filtering) SIMO system with channel coefficients $h_i$, transmit power $\kappa_p$, 
and noise variance $\sigma_n^2$. 
Assuming a zero-mean complex circular symmetric Gaussian channel such that  $\mathbf{h} \sim  {\cal C}{\cal N}(0, \sigma_h^2 I)$, 
the scaling law of the expected value of $P_e^u(\varsigma_h, \epsilon_p)$ with respect to the channel is 
\begin{eqnarray}
\log E_{\varsigma_h} \left( P_e^u(\varsigma_h, \epsilon_p) \right)
 \approx  M \log \left( \frac{\rho  {{\kappa_p}}}{2} \right), 
\label{PePerChUPInstRayleighCh}
\end{eqnarray}
where the average SNR over the channel energy is defined as $\rho = \frac{\sigma_h^2}{\sigma_n^2}$.

In general, the SER depends on the DoF contained in $\varsigma_h$.
Consider a  sparse channel with $L< M$ paths and a simple case where all the paths have equal energy and are separated by a directional cosine angle larger than one over the array aperture length \cite{Tse2004}, we have $\varsigma_h \approxeq \sum_{l = 0}^{L-1} \vert \beta_l \vert^2$, where 
$\beta_l$ is the complex gain of the $l$th path. Then, we can prove that  
\begin{eqnarray}
\log P_e^u(\varsigma_h, \epsilon_p) 
 \approx  L \log \left(\frac{\rho \kappa_p M}{2L} \right). 
\label{PePerChUPInstSparseCh}
\end{eqnarray}
Clearly,  compared to \eqref{PePerChUPInstRayleighCh}, when $L < M$, the performance in \eqref{PePerChUPInstSparseCh} is limited by the reduced DoF $L$.

\subsubsection{Comparison with Coherent Detection}\label{Sec:Cohe}
For coherent detection, the received signal after matched filtering and normalization is
\begin{equation}
	z_c = \sqrt{\epsilon_p} + \frac{\mathbf{h}^H \mathbf{n}}{ \vert\vert \mathbf{h} \vert\vert^2_2}.
\label{eq:CohD}
\end{equation}
Detection of $\sqrt{\epsilon_p}$ is based on the real part of $z_c$ as the imaginary part does not contain any information about $\sqrt{\epsilon_p}$. Based on (\ref{eq:CohD}), the detection threshold is  $\Delta_p = \frac{\sqrt{\epsilon_p} + \sqrt{\epsilon_{p+1}}}{2}$. 
It is straightforward to determine the corresponding post-processing SNRs $\gamma_{u,p}$ and $\gamma_{l,p}$ and prove that they are equal to the expression in (\ref{GammaRelations}), see  Appendix \ref{APPdix:OptimalConst}. Therefore, the performance of  I-ED asymptotically approaches the performance with coherent receivers. 

\subsection{Constellation Design}\label{Sec:InstconstOpt}

Inspecting \eqref{PePerChUPInst}, \eqref{PePerChLowTailInst} and \eqref{PeInst}, the performance depends on the power of the constellation points. Hence, for a given constellation size, it is possible to design the power distribution of the constellation points to optimize the SER. Optimizing the average SER is a sensible approach when the channel holds a large number of DoFs. Here, we optimize the instantaneous SER, so that the design also holds for channels with a limited number of DoFs. 

The  optimization problem is formulated as:
\begin{align} \label{OptInst}
& \arg\min_{\epsilon_0, \ldots, \epsilon_{P-1}} P_e(\varsigma_h) \\
&  \text{Subject to  } \frac{1}{P}\sum_{p = 0}^{P-1} \epsilon_p = 1, \notag \\      
& \hspace{50pt}\epsilon_i < \epsilon_j \; \text{when} \;\; i<j. \notag 
\end{align}
In general, this is a NP-hard problem.  Similarly to \cite{Hammouda2015}, 
the convexity properties of the Q function with respect to the threshold values can be exploited leading to an iterative algorithm with low complexity.

Employing the threshold values in \eqref{PAMdeltaInst} and the asymptotic moments in \eqref{PAMMomentsInst}, we can in fact compute the optimal constellation at high SNR regimes in closed-form. We show in Appendix \ref{APPdix:OptimalConst} that the optimal constellation, achieving the lowest SER, is the conventional non-negative PAM. 
In this case, $\sqrt{\epsilon_{p+1}} - \sqrt{\epsilon_{p}}$ takes the same value for $p = 0, \ldots, P-2$, leading to equal 
$\gamma_{l,p}$ and $\gamma_{u,p}$ at high SNR regimes. 
Though no-closed form solution can be obtained at low and medium SNRs, an alternative design criteria could be to equalize the post-processing SNRs, which can be formulated as a minimax problem
\begin{align} \label{OptInst1}
& \arg\min_{\epsilon_0, \ldots, \epsilon_{P-1}} \max_{p} \vert \gamma_{l,p} - \gamma_{u,p}\vert^2 \\
&  \text{Subject to  } \frac{1}{P}\sum_{p = 0}^{P-1} \epsilon_p = 1. \notag   
\end{align}
It can be efficiently be solved by standard optimization toolboxes.

\section{Average channel energy based ED (A-ED)}
\label{sec:PA}

In this section, ED is performed based on the knowledge of the average channel energy 
$\sigma_h^2$ and the noise energy $\sigma_n^2$ which are assumed to be perfectly known. 
A-ED is suited for fast fading channels where the receiver cannot track the instantaneous channel energy, but also for channels holding a large number of DoFs. 

Next, we determine the detection thresholds for A-ED based on two methods. The first method relies on a Gaussian approximation of the received signal after energy collection while the second approach uses a Bayesian approach. Then, we analyze the average SER which is expressed simply as a function of Chi-square cdfs. This allows for a simple formulation of  input constellation optimization based on the average SER. 

\subsection{Decision Thresholds}
\subsubsection{Gaussian Approximation}\label{Approach2}
In slow or time-invariant fading channels, using training symbols to estimate the instantaneous channel energy $\varsigma_h$ justifies the application of \eqref{zMean} and \eqref{zVar}. However, in fast fading channels in which $\sigma^2_h$ stays constant, ED based on an estimate of $\varsigma_h$ provides inferior performance due to the fact that the instantaneous channel energy in the training and data transmission phases may be different. Thus a larger estimation variance is obtained compared to employing average channel energy. In this case, we can directly reuse the results in Section \ref{Approach1} by approximating $ \varsigma_h$ with $\sigma^2_h$.

\subsubsection{Bayesian Energy Detection Using a Priori Knowledge of Channel Distribution} \label{App2}

We now exploit the prior distribution of $\bfh$ which is $\bfh \sim {\cal C }{\cal N}(0, \sigma_h^2)$. 
Inspecting \eqref{EDoutput} and employing the assumption on $\bfh$,
$z$ follows a centered Chi-square distribution with pdf 
\begin{equation}\label{eq:GauPrioChiSquare}
\begin{array}{l}
f_b (z|\sigma^2_h,\epsilon_p) = 
\left(\frac{M}{ \sigma^2_h \,\epsilon_p + \sigma_n^2} \right)^M
e^{  -\frac{M}{\sigma^2_h \,\epsilon_p+ \sigma_n^2} z}
 \frac{ z^{M-1}}{(M-1)!} \; \text{for } z >0 .
\end{array}
\end{equation}
Note that we use the subscript ``$b$'' to indicate the conditional pdf of $z$ is obtained via a Bayesian approach.

Applying~\eqref{eq:GauPrioChiSquare} to \eqref{eq:MLD}, we find that the detection thresholds have a closed-form expression:
\begin{align}
\Delta_p & =
\left[\log \left(
\frac{ \rho  \epsilon_{p+1} + 1}
{ \rho \epsilon_{p} +  1  }
\right) 
+ \frac{1}{M} \log \left(\frac{p(\epsilon_{p})}{p(\epsilon_{p+1})}  \right)\right] \notag \\
& \cdot
\frac{( \rho\epsilon_{p+1} +  1 )( \rho \epsilon_{p} +  1 )}
{\rho (\epsilon_{p+1} - \epsilon_{p})}.
\label{eq_Thr3}
\end{align}
If $M$ is sufficiently large and the transmit symbols are equi-probable,~\eqref{eq_Thr3} simplifies to the results given in \cite{Martinez2014}:
\begin{align}\label{ThG}
\Delta_p & =
\log \left[
\frac{( \rho  \epsilon_{p+1} +  1 )}
{( \rho \epsilon_{p} +  1 ) }
\right]\frac{( \rho\epsilon_{p+1} +  1 )( \rho \epsilon_{p} +  1 )}
{\rho (\epsilon_{p+1}- \epsilon_{p})}.
\end{align}
Remarkably,  
for a large number of antennas $M$, the threshold value does not depend on $M$ but only on the SNR.

\subsection{Performance Analysis}

Employing the decision thresholds, we now determine the  expected value of the SER with respect to the channel distribution. 
Our main assumption is that $\Delta_p$ is computed in the training phase using the average channel energy and thus does not depend on the channel $\bfh$ in the data transmission phase. In a fast fading channel, this assumption is verified when the channel in the training phase is independent from that of the data transmission phase. 
If the training in the current time slot is incorporated, the estimate of $\sigma_h^2$ depends on the channel in the current time slot. To remove this dependency, we can simply ignore the training in this slot. Alternatively, if the number of training slots is sufficient such that the correlation of the estimate with the channel in the current time slot is negligible, our results in the following still hold. 

Focusing on $P_e^u(\varsigma_h, \epsilon_p)$ defined in (\ref{PeUInst}), we have
\begin{align}
E_{\varsigma_h}[P_e^u(\varsigma_h, \epsilon_p)] & =  
\int_{\varsigma_h} \int_{z=\Delta_p}^{\infty}   f(z| \varsigma_h,\epsilon_p)  dz  f(\varsigma_h) \mathrm{d}\varsigma_h \label{IntSwapG}.
\end{align}
Based on the independence assumption between $\Delta_p$ and $\varsigma_h$ in the data transmission phase, we  swap the integrals in~\eqref{IntSwapG} to compute the average SER. Then, we use the fact that $\int_{\varsigma_h}  f(z|\varsigma_h, \epsilon_p) f(\varsigma_h) d\varsigma_h = f(z| \epsilon_p)$ which is a Chi-square pdf with known asymptotic properties. Accordingly, we obtain the result: 
\begin{align}
E_{\varsigma_h}[P_e^u(\varsigma_h, \epsilon_p)]  =
\int_{\Delta_p}^{\infty}   f(z| \epsilon_p)  dz \label{Peu} 
& = 1 - F_{z} (\Delta_p|\epsilon_p),
\end{align}
which can be completely characterized by a Chi-square cdf. Similarly, the probability  of decoding $\epsilon_{p-1}$ while $\epsilon_{p}$ is transmitted reads 
\begin{align}\label{Pel}
E_{\varsigma_h}[P_e^l(\varsigma_h, \epsilon_{p})] & = 
F_{z} (\Delta_{p-1}|\epsilon_{p}).
\end{align}

Using~\eqref{Peu} and ~\eqref{Pel}, we obtain closed-form expressions for the SER of $\epsilon_p$:
\begin{equation} \label{PeAvg}
\begin{array}{l}
P_e(\epsilon_p) =
\left\{ \hspace{-2mm}
\begin{array}{lr}
1- F_{z}( \Delta_p|\epsilon_p),  &  p=0 \\
F_{z}( \Delta_{p-1}| \epsilon_p) + 1\! -\! F_{z}( \Delta_{p}| \epsilon_p),  & 
\hspace{-1.5mm}  0<p<P\!-\!1  \\
F_{z}( \Delta_{p-1}| \epsilon_p), &   p=P\!-\!1.
\end{array}
\right.
\end{array}
\end{equation}
Note that the SER is fully characterized by Chi-square cdfs with $2M$ DoFs. This is distinguishable with the results presented in~\cite{Martinez2014}, where the average SER needs to be obtained via Monte Carlo simulations. Thus, the average SER over the transmit symbols is given by 
\begin{equation}\label{Pe}
P_e = \sum_{p = 0}^{P-1} P_e(\epsilon_p) p(\epsilon_p).
\end{equation}

\subsection{High SNR Analysis}

To provide the scaling law behavior of $P_e$, we use the Chernoff bounds to approximate the tails of a Chi-square cdf~\cite{J.A.Gubner2006}. It gives a better approximation of the tails of the distribution than the known first or second moment based tail bounds such as Markov's inequality or Chebyshev inequality~\cite{J.A.Gubner2006}. To apply this bound, for the problem at hand, we define 
\begin{equation}\label{pu}
\delta_p^u= \frac{\Delta_p}{\sigma^2_h\, \epsilon_p  + \sigma_n^2} \;\;\; \text{for} \; p = 0,\ldots,P-2
\end{equation}
\begin{equation}\label{pl}
\delta_p^l = \frac{\Delta_{p-1}}{\sigma^2_h\, \epsilon_p  + \sigma_n^2} \;\;\; \text{for} \; p = 1,\ldots,P-1.
\end{equation}
Therefore, an approximation of~\eqref{Peu} reads
\begin{equation}\label{PeuA}
P_e^u(\sigma^{2}_h , \epsilon_p) \approx \left[ \delta_p^u e^{1- \delta_p^u} \right]^M.
\end{equation}
Likewise, the approximation of the lower tail of the Chi-square distribution in~\eqref{Pel} has the form 
\begin{eqnarray}\label{PelA}
P_e^l (\sigma^{2}_h, \epsilon_p) \approx \left[ \delta_p^l e^{1- \delta_p^l} \right] ^M.
\end{eqnarray}
Having~\eqref{PeuA} and~\eqref{PelA}, the approximate SER can be readily obtained. The application of this bound sheds some light on the performance of the adopted constellation schemes.

\subsubsection{OOK Constellation}\label{OOKex}
In this case, the energy constellations are denoted as $\epsilon_0 = 0$ and $\epsilon_1 = 2$ such that the average power is unity. To analyze the SER performance, we define $ \rho'_1 =  \epsilon_1 \rho$. The detection thresholds are determined using~\eqref{eq:roots} as well as 
\eqref{zMean} and \eqref{zVar}. 
When  $\rho$ $ \rightarrow +\infty$, the detection thresholds are approximated as
\begin{equation}
\Delta_0 =  \sigma_n \sqrt{\frac{ \sigma^{2}_h \,\epsilon_1 }{2}} =  
\sigma_n^2 \sqrt{\frac{\rho'_1 }{2}}. 
\end{equation}
Applying~\eqref{pu} and~\eqref{pl}, the parameters controlling the upper tail of $p(z|\sigma^2_h,\epsilon_0)$ and the lower tail of $p(z|\sigma^2_h,\epsilon_1)$ are given by
\begin{equation}
\delta_0^u =  \sqrt{\frac{ \sigma^2_h \,\epsilon_1   }{2\sigma_n^2}} =  \sqrt{\frac{  \rho'_1}{2}}  \quad \mbox{and}\quad
\delta_1^l = \frac{\sigma_n}{\sqrt{2   \sigma^2_h\,\epsilon_1    }}= \frac{1}{\sqrt{2 \rho'_1 }}.
\end{equation}
Inserting these two terms in~\eqref{PeuA} and~\eqref{PelA}, it can be shown that $P_e( \epsilon_1) \gg P_e(\epsilon_0)$: see Appendix \ref{App1}. Thus, the average SER is dominated by $P_e(\epsilon_1)$. Taking the logarithm of the SER, we obtain   
\begin{eqnarray}
\log P_e \approx \log P_e(\epsilon_1) \approx  M  \log \delta^l_1 =  - \frac{M}{2}  \log 2\rho'_1.
\end{eqnarray}
It is found that using OOK, the ED achieves diversity order of $M/2$. Although lower than the diversity order achieved by coherent detection, equal to $M$, 
the proposed receiver obtains a diversity order that linearly scales with the number of antennas. The loss in diversity is due to the absence of knowledge of the instantaneous energy and the fact that the receiver relies on the average energy.
 
\subsubsection{Higher-order PAM Constellation} \label{PAMSERanalysis}
For a general PAM constellation, when  $\rho$ $ \rightarrow +\infty$, using~\eqref{eq:roots}, \eqref{zMean}, and \eqref{zVar}, the detection thresholds are approximated as  
\begin{equation}\label{PAMdelta}
\begin{aligned}
\Delta_0 &=   
\sigma_n^2 \sqrt{\frac{\rho'_1 }{2}}  \\
\Delta_p & =  \sigma^2_h \sqrt{\epsilon_p} \sqrt{\epsilon_{p+1}}, \;\;  \text{for} \;p\neq 0. 
\end{aligned}
\end{equation}
Inserting \eqref{PAMdelta} to~\eqref{pu} and~\eqref{pl}, the parameter controlling the lower and upper tails are given by
\begin{align} \label{PAMTail}
\delta_0^u &=  \sqrt{\frac{ \sigma^2_h \,\epsilon_1 }{2\sigma_n^2}} =  \sqrt{\frac{  \rho'_0}{2}} \\
\delta_p^u &= \frac{ \sqrt{\epsilon_{p+1}}}{ \sqrt{\epsilon_{p}}} \left(\frac{\rho'_p}{\rho'_p + 1}\right) \;\;\; \text{for} \; p = 1,\ldots,P-2 \\
\delta_p^l &= \frac{ \sqrt{\epsilon_{p-1}}}{ \sqrt{\epsilon_{p}}} \left(\frac{\rho'_p}{\rho'_p + 1}\right) \;\;\; \text{for} \; p = 1,\ldots,P-1,
\end{align}
where $ \rho'_p = \epsilon_p \rho$. Applying these three terms in~\eqref{PeuA} and~\eqref{PelA}, it can be shown that the  SER is given by   
\begin{eqnarray}\label{eq:PAMPe}
P_e \approx \sum_{p = 1}^{P-2} \left(\left( \eta_{p} e^{(1 - \eta_{p})}\right)^M + \left(\frac{1}{\eta_{p}} e^{(1 - \frac{1}{\eta_{p}})}\right)^M\right) p(\epsilon_p),
\end{eqnarray}
where $\eta_{p} = \sqrt{\frac{\epsilon_{p+1}}{ \epsilon_{p}}}$. We immediately observe that \eqref{eq:PAMPe} depends only on the number of receiver antennas and the power ratio between the constellation points. Thus, for fixed $M$, the SER in \eqref{eq:PAMPe} tends to a constant, leading to an error floor, which depends on the energy level of the signal constellation. 

For the conventional non-negative PAM, the values of the first and second term in the summand in \eqref{eq:PAMPe} are more and more similar as the constellation order $P$ increases. When $P$ is small, e.g. $P = 4$ with an uniform distributed constellation, it can be shown that the SER is dominated by the term $p = 2$ in \eqref{eq:PAMPe}:
\begin{equation}\label{eq:PAMPe4PAM}
P_e \approx \left( \eta_{2} e^{(1 - \eta_{2})}\right)^M + \left(\frac{1}{\eta_{2}} e^{(1 - \frac{1}{\eta_{2}})}\right)^M.
\end{equation}
in which the second term is more significant. Thus, we obtain
\begin{eqnarray}\label{eq:PAMPeApprox}
\log P_e \approx -\frac{M}{2} \log \left( \eta^2_{2}  + constant\right).
\end{eqnarray}
Therefore, the logarithm of the SER decreases linearly with the number of antennas and it is beneficial to employ more antennas since it leads to a reduced probability of error. 
 
\subsection{Constellation Design} \label{Sec:CDdesign}

Based on the average channel energy, we can optimize the constellation. 
The optimization problem can be formulated in a similar way as \eqref{OptInst} but with the SER expression in \eqref{Pe}. To compute the solution, the iterative optimization algorithm in \cite{Hammouda2015} can be employed for our proposed optimization problem. The Chi-square cdf in \eqref{PeAvg} has convexity properties with respect to the corresponding threshold values, similarly to the Q function in \cite{Hammouda2015}. 
Thus, the method can be straightforwardly applied to our problem with only a few iterations required to obtain a sub-optimal solution.

\section{Performance Assessment and Simulations}
\label{sec:sim}

We report the performance of the proposed ED methods using OOK and $4$-PAM. In fast fading channels, the estimated channel energy in the training phase may be largely different from that of the data transmission phase. If this occurs, the use of instantaneous channel energy leads to an inferior performance when I-ED or coherent receiver is utilized: see the performance evaluation in \cite{Martinez2014}. In slow fading channels, the instantaneous channel energy stays the same in the training and data transmission phase, which allows a meaningful comparison between coherent and non-coherent methods. We thus only report the results in slow fading channels. In addition, for $4$-PAM constellations, we compare the performance of ED using the conventional $4$-PAM constellation and the optimized PAM constellation. To compare with the methods in the literature, i.e. \cite{Hammouda2015}, we assume that the energy levels have an uniform probability mass function. 
 Defining the term SNR as  $\rho = \frac{\sigma^2_h}{\sigma^2_n}$ in this section, we compare the SER obtained for the following cases: 
\begin{itemize}      
	\item  ``Coherent'': Coherent detection with known CSI at the receiver. The SER is computed by averaging over Monte Carlo trials of the channel.
	\item ``I-ED: ChiS'': ED based on instantaneous channel energy $\frac{1}{M}\sum_{i=1}^M |h_i|^2$ and $\sigma_n^2$. The SER is computed using the upper tail \eqref{PeUInst} and the corresponding lower tail based on the exact (Chi-square) pdf of the received signal energy. 
	\item ``I-ED: Gaus'': ED based on instantaneous channel energy  $\frac{1}{M}\sum_{i=1}^M |h_i|^2$ and $\sigma_n^2$. The SER is computed using \eqref{PePerChUPInst} and \eqref{PePerChLowTailInst}  based on the Gaussian approximation of the received energy in Section \ref{Approach1} and \ref{Sec:BF}.  
	\item ``A-ED-Gaus'': ED based on average channel energy  $\sigma_h^2$, $\sigma_n^2$ and the Gaussian approximation of the received energy in Section \ref{Approach2}. The analytical SER is obtained from \eqref{PeAvg}.
	\item ``A-ED-Bayesian'': ED based on $\sigma_n^2$ and the prior knowledge of the channel in Section \ref{App2} to compute the thresholds. The analytical SER is obtained from \eqref{PeAvg}.
	\item ``I(A)-ED: Opt-PAM'': ED  using the optimized constellation based on the method ``I(A)-ED: Gaus''. 
	\item ``A-ED: (Opt-PAM) Gaus-\cite{Hammouda2015}'': Analytical SER obtained from using (optimized constellation design) conventional PAM in \cite{Hammouda2015} based on Gaussian approximations.
	\item ``I-ED: SparseCh: EqualPow'': I-ED for a sparse channel with equal power path  gain.
	\item ``I-ED: SparseCh: UnEqualPow'': I-ED for a  sparse channel with exponential decay power delay spectrum.
\end{itemize}
%
%

\begin{figure}
	\centering
    \resizebox{0.9\columnwidth}{!}{\includegraphics{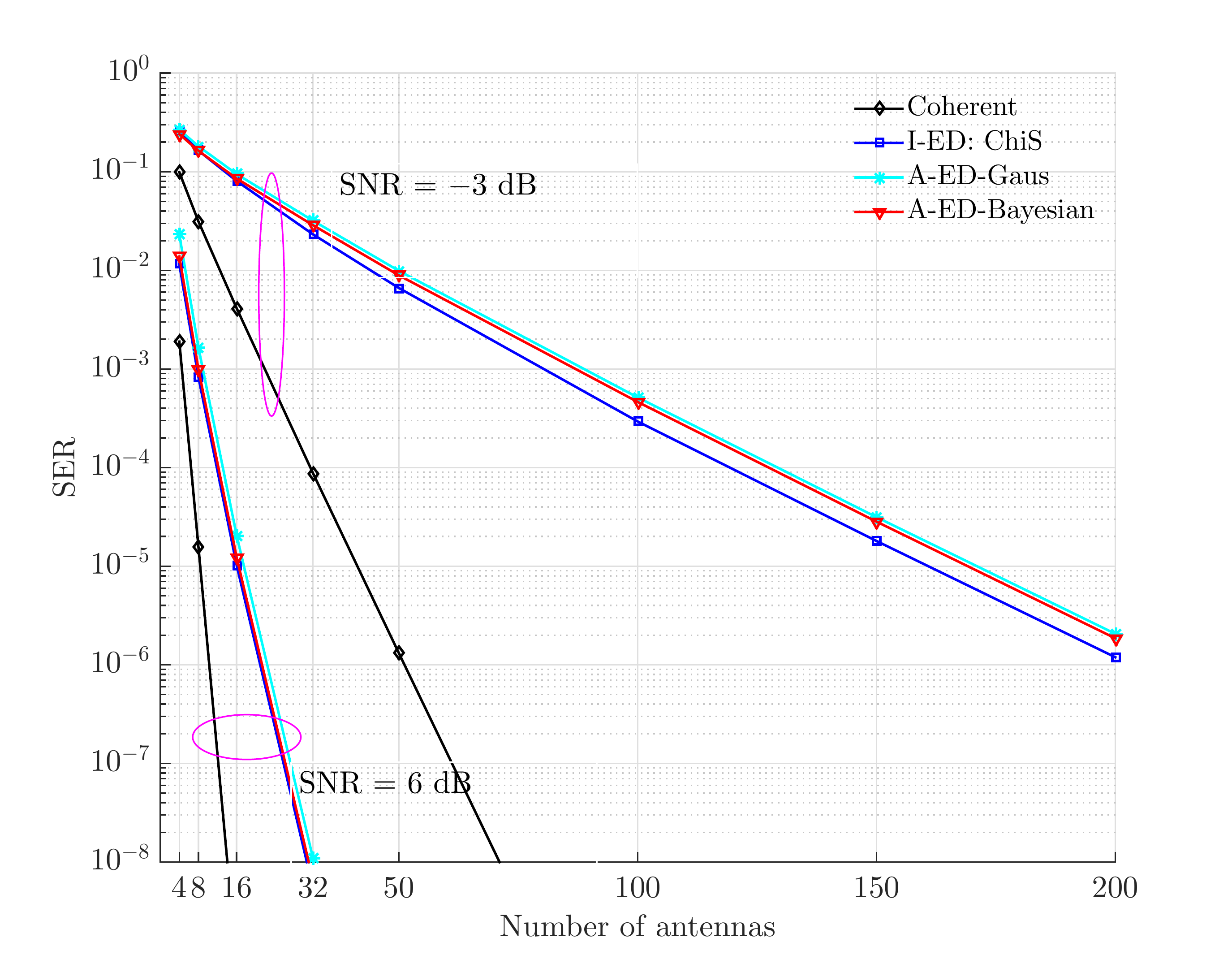}}
	\caption{OOK: SER versus the number of antennas.}
	\label{OOK}                              
\end{figure}

\subsection{OOK Constellation}


Fig. \ref{OOK} shows the SER performance using the OOK constellation. We observe that the performance gap between coherent detection and the proposed ED methods becomes smaller and smaller as the SNR increases. In fact, when I-ED is employed, its performance asymptotically approaches that of  coherent detection: see Section \ref{Sec:Cohe}. As the number of antenna increases, the SER decreases, which shows the benefits of employing a large number of antennas in the ED system. As an example, inspecting the red curves, to achieve SER $ = 10^{-3}$, employing $M = 100$ antennas leads to around $\SI{9}{\dB}$ gain compared to using $M = 8$, corresponding to a reach extension of roughly $\SI{2.8}{\km}$\footnote{The standard path loss model reads: PathLoss [$\si{\dB}$] = $20\log_{10}(d)$, where $d$ is the reach extension with unit $\si{\km}$. Note that for indoor applications, the distance $d$ has unit in meter.}.  In addition, no error floor is observed as the SNR increases, which is analyzed in Section \ref{OOKex}.

\subsection{$4$-PAM}
\begin{figure}
	\centering
	\resizebox{0.9\columnwidth}{!}{\includegraphics{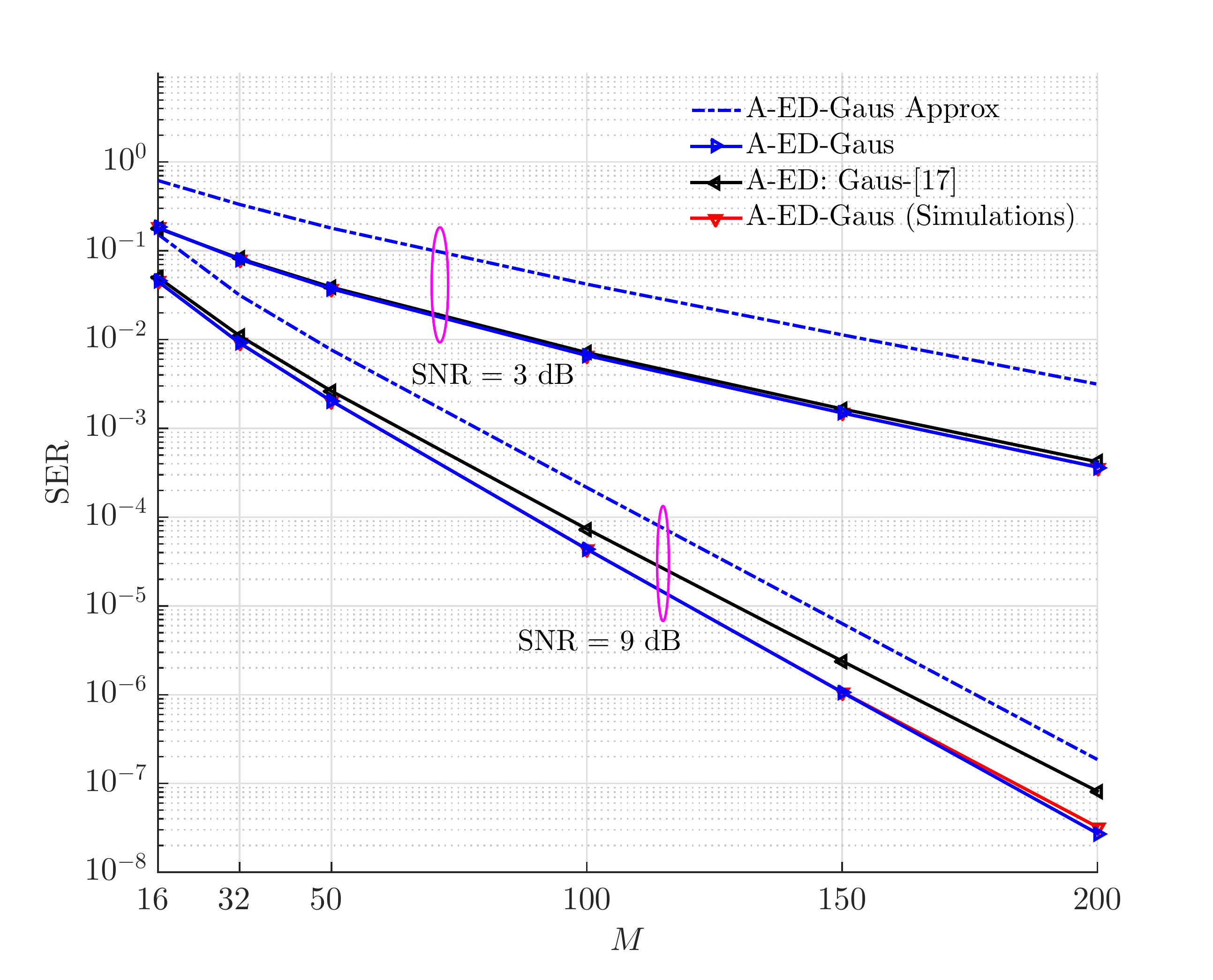}}
	\caption{{Conventional $4$-PAM: Comparison of the proposed analytical SER expression in \eqref{PeAvg}, the approximation using~\eqref{PeuA} and~\eqref{PelA}, denoted as ``A-ED-Gaus Approx'', and the solution in \cite{Hammouda2015}. }
}
	\label{4PAM_compare}                              
\end{figure}

\begin{figure}
	\centering
	\resizebox{0.9\columnwidth}{!}{\includegraphics{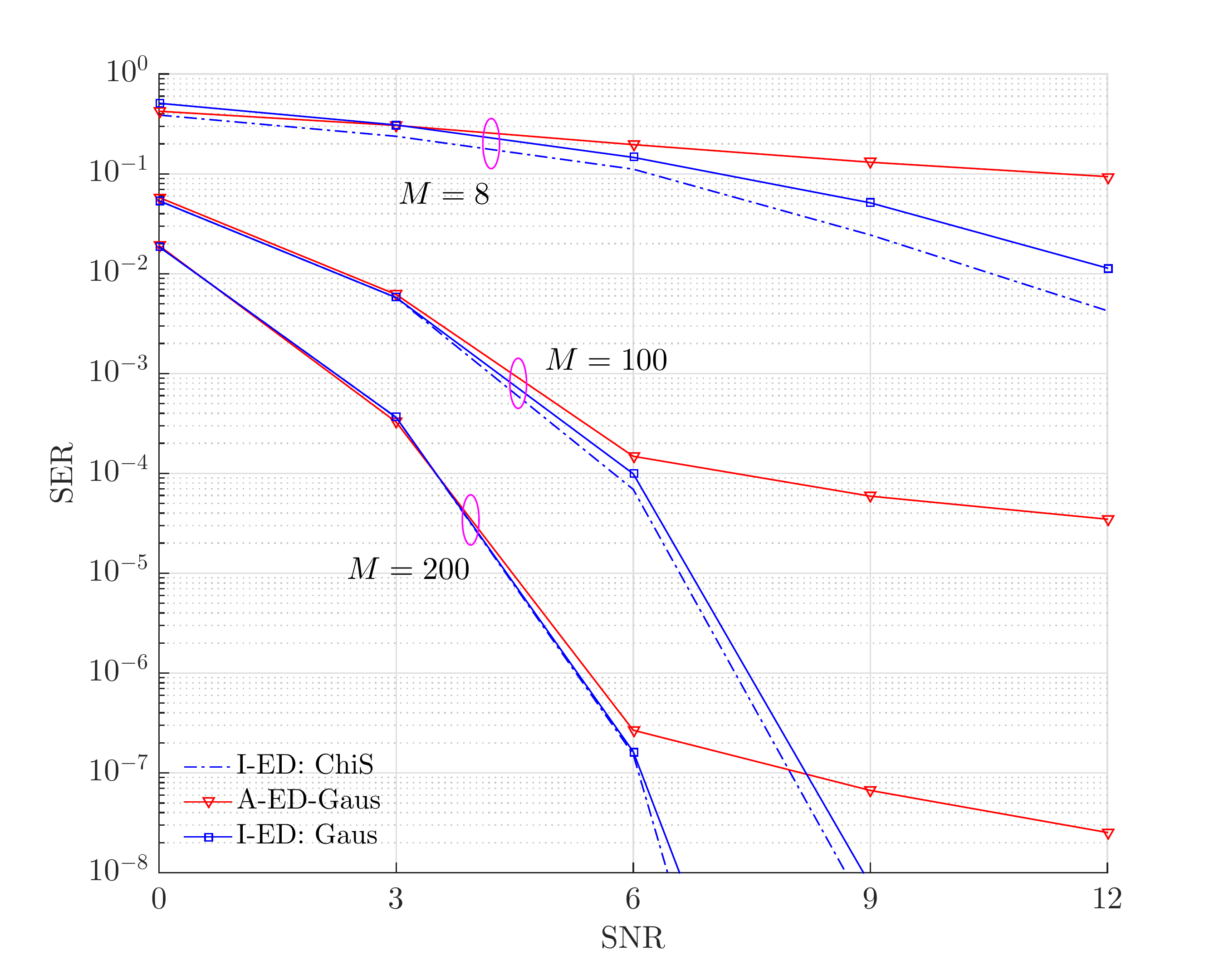}}
	\caption{Conventional $4$-PAM: SER versus SNR for different No. of antennas.}
	\label{4PAMsnr}                              
\end{figure}

Assuming conventional PAM is adopted at the transmitter, Fig. \ref{4PAM_compare} compares the proposed SER expression using the exact Chi-square pdfs to the Gaussian approximation based-solution in \cite{Hammouda2015}. The proposed solution \eqref{PeAvg} fits the Monte Carlo simulation rather well at both low and high SNR regimes while a noticeable difference between the two approaches is observed at high SNRs, which is caused by the adopted Gaussian approximations in \cite{Hammouda2015}. {In addition, we also observe that the approximations~\eqref{PeuA} and~\eqref{PelA} can be used to predict the slope of the SER, which justifies its application to analyze the behavior of the SER, especially at high SNR regimes where the error floor appears, see Fig. \ref{4PAMsnr}. This error floor is caused by the fact that the SER performance is limited by the uncertainty in the channel energy. 
As it is analyzed in Section \ref{PAMSERanalysis}, we can see that increasing the number of antennas reduces the error floor, but does not remove it. When the instantaneous channel energy is known, as the number of antenna increases, we observe a diminishing performance gap resulting from the Gaussian approximation of the pdfs to compute the SER as described in Section \ref{Approach1} compared to the exact SER based on the 
Chi-square pdfs.  }

\begin{figure}
	\centering
	\resizebox{0.9\columnwidth}{!}{\includegraphics{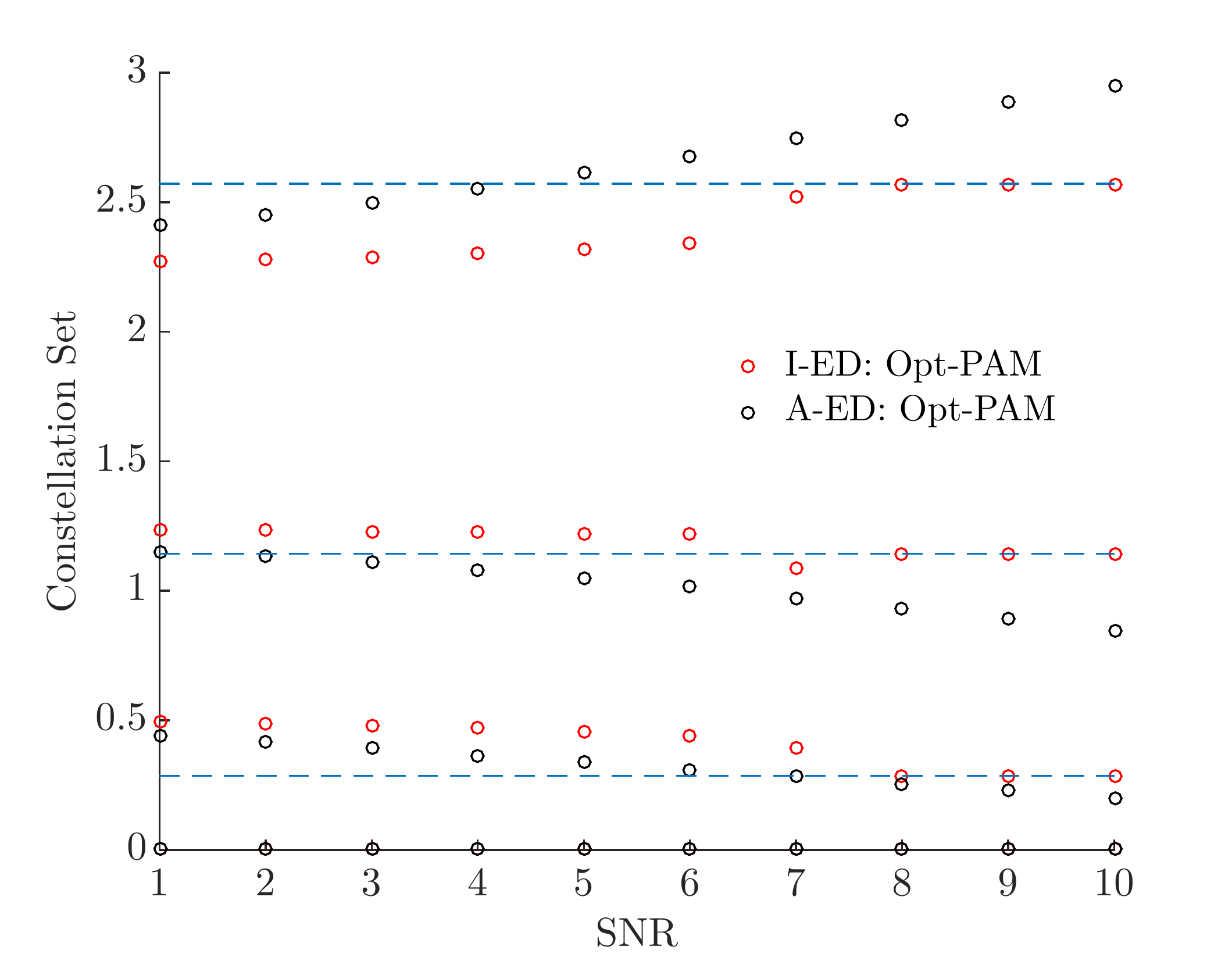}}
	\caption{$4$-PAM energy constellation optimization: $M = 100$. The dashed line shows the energy level of the conventional PAM.}
	\label{Constopt}                              
\end{figure}

In Fig. \ref{Constopt}, we compare I-ED and A-ED when the input constellation is optimized using the methods described in Section \ref{Sec:InstconstOpt} and \ref{Sec:CDdesign},  respectively. 
For A-ED  and as the SNR increases, the energy of the symbol with the highest energy becomes larger and larger, while the distance between the other three gets smaller and smaller. This coincides with the analysis in \eqref{eq:PAMPe} where the SER is mainly dominated by the energy ratio between the two largest energy constellations. 
For I-ED, as the SNR increases, the optimal constellation converges to the  conventional PAM.
This verifies our analysis in Section \ref{BF:analysis} which shows that the performance of I-ED approaches the performance of the coherent receiver as the SNR increases. Since conventional PAM is optimal in terms of achieving the lowest SER for a coherent receiver, it becomes 
the optimal constellation for I-ED at high SNR regimes. 
 
\begin{figure}
	\centering
	\resizebox{0.9\columnwidth}{!}{\includegraphics{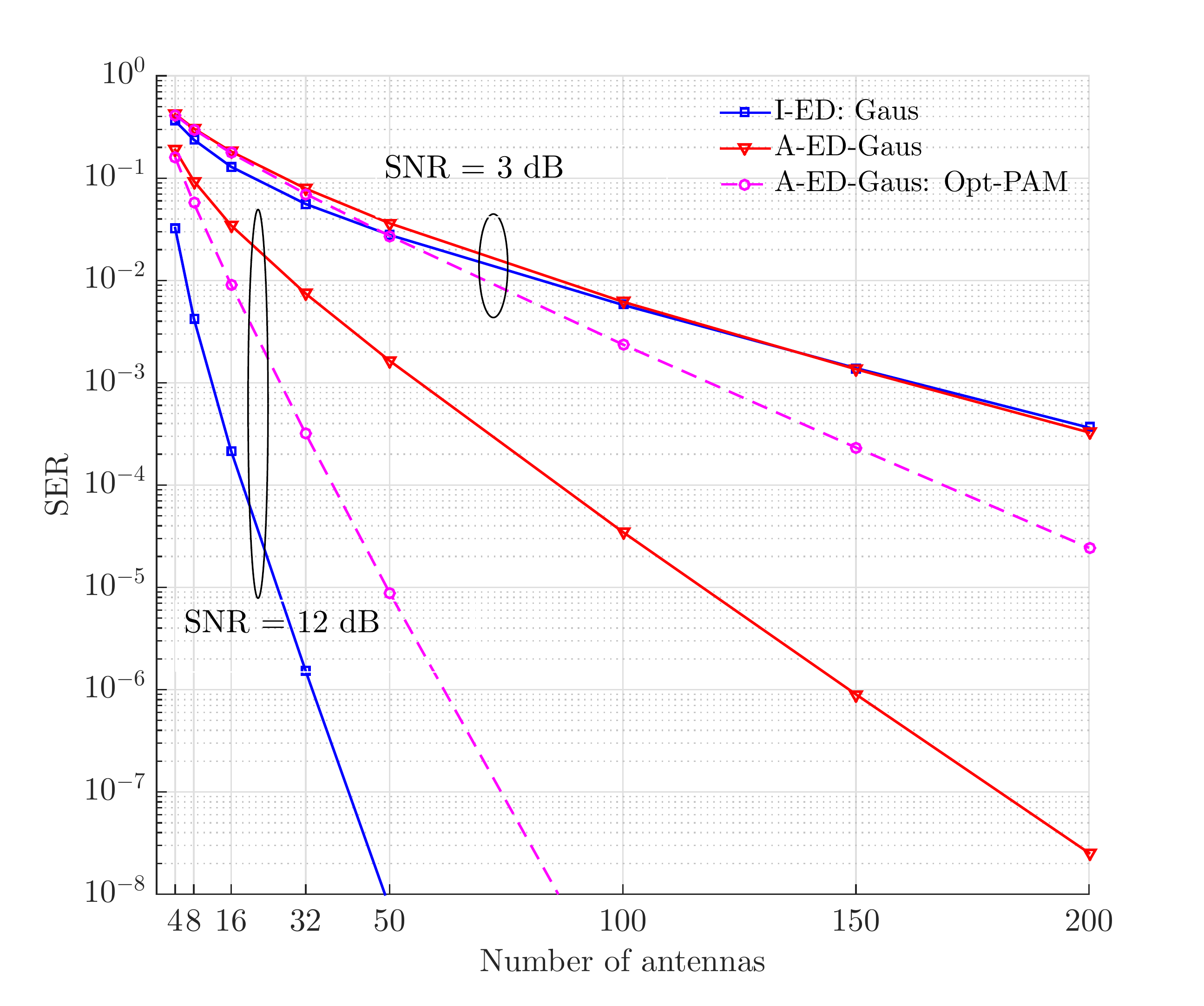}}
	\caption{Constellation optimization: SER versus the no. of antennas.}
	\label{SERopt}                              
\end{figure}

Fig. \ref{SERopt} reports the performance of A-ED with optimized PAM for a $4$-level constellation. 
The results show that using an optimized energy constellation can significantly improve the SER in the considered SNR regions. In addition, the error floor is not observed and appears at very high SNR. Thus, A-ED together with an optimized PAM constellation brings appealing performance provided that the channel holds a very large number of DoF.

\subsection{Performance Evaluation in Sparse Channels}

In Rayleigh fading channels, A-ED delivers promising performance as it is reported in the previous sections. However, the performance of A-ED significantly degrades when the number of DoF becomes small, e.g. in sparse channels. 

To evaluate the ED methods, we adopt the sparse channel model from \cite{Bajwa2010}:
\begin{equation}\label{SparseCh}
\mathbf{h} = \underbrace{\beta_{0} \mathbf{v}(\theta_{0})}_{\mathrm{LOS\; component}} + \underbrace{\sum_{l=1}^{L-1} \beta_{l} \mathbf{v}(\theta_{l})}_{\mathrm{Non-LOS (NLOS) \; component}},
\end{equation}
where $L$ is the number of paths, $\beta_{l} \sim \mathcal{CN}(0,\sigma^2_{\beta_{l}})$  and $\theta_{l}$, for $l = 1, \ldots, L-1$,  are the complex amplitude and the arriving angle of the $l$th path. For simplicity, we set $\beta_{0}$ to a positive constant for LOS channels and $\beta_{0}=0$ for NLOS propagation environment. In mmWave systems, $L$ is typically small \cite{Rappaport2013}. The LOS component arrives at the receiver array from the direction $\theta_{0}$ if it exists. Without loss of generality, we assume that the channel power is normalized. The steering vector is defined as $\mathbf{v}(\theta_{l})  = [1, \ldots, e^{-j \pi (M-1)  \cos(\theta_{l})}]$ assuming a half wavelength spacing between the antenna elements and $j = \sqrt{-1}$. To simplify, we assume that the paths have fixed angle of arrivals with uniform spacing in $[0, 2\pi]$, so that the inner product of the steering vectors tends to zero as the number of antennas tends to infinity. When $L\leq M$, each path occupies a separable beam and the DoF is limited by $L$. When $L > M$, the maximum DoF is determined by $M$, which means that some paths become unresolvable given the array aperture. We conduct Monte-Carlo simulations to report the performance of the ED methods under the channel model \eqref{SparseCh} using error expressions such as \eqref{PeUInst}.

In LOS channels with a Rician factor, defined as $\frac{\beta^2_{0}}{\sum_{l=1}^{L-1} \sigma^2_{\beta_{l}}}$, set to be $\SI{9}{\dB}$, 
we compare the performance of the Opt-PAM and Conv.-PAM assuming that the NLOS paths have equal power. We observe that the optimized constellations result in noticeable performance improvement, especially the design based on instantaneous channel energy. Meanwhile, our proposed designs outperform that from \cite{Hammouda2015}.

\begin{figure}
	\centering
	\resizebox{0.9\columnwidth}{!}{\includegraphics{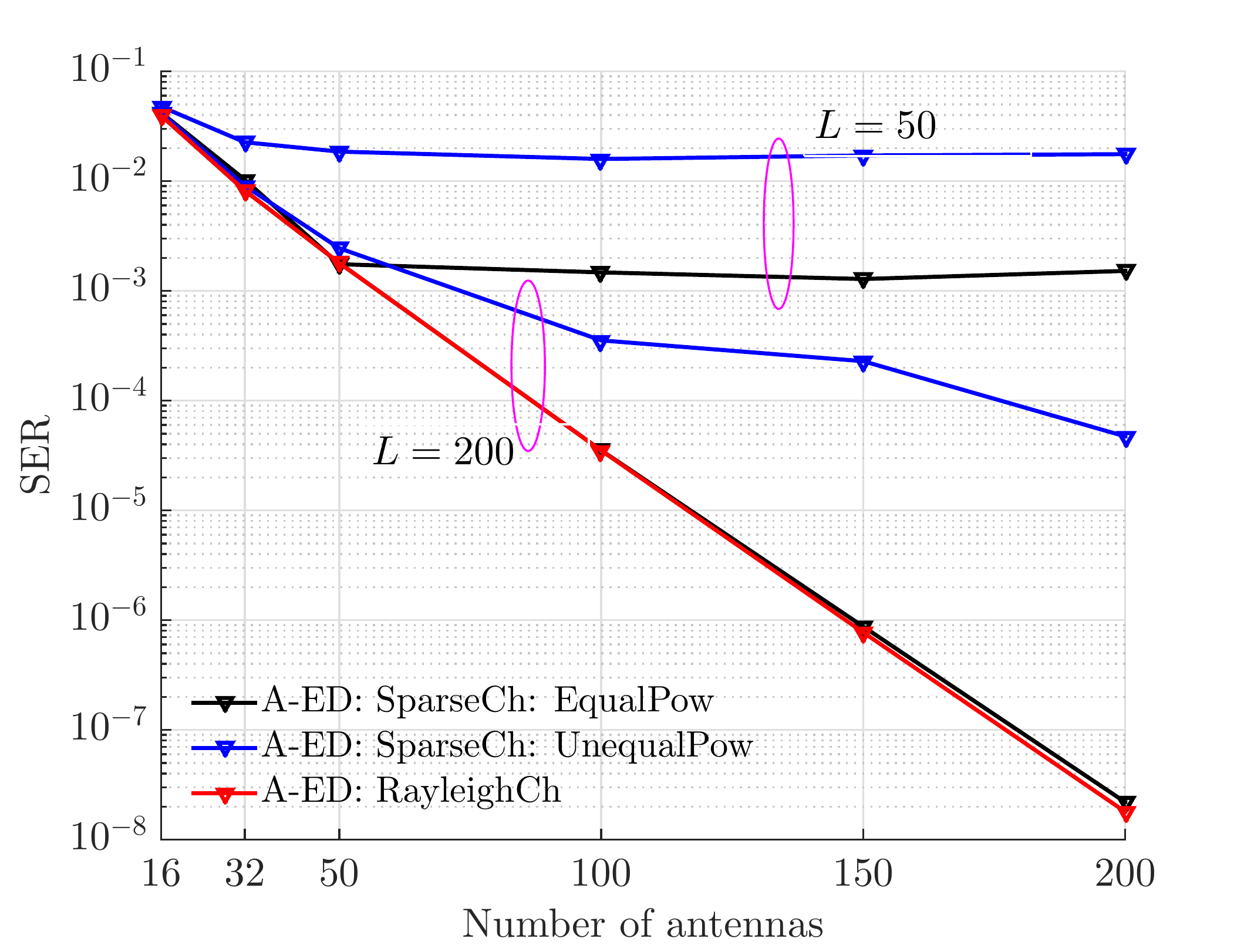}}
	\caption{SER versus the no. of antennas for different numbers of path components in NLOS sparse channels: Conventional PAM, SNR $= \SI{10}{\dB}$ and using average channel energy in Section \ref{App2} to compute the threshold values.}
	\label{SERFourier_Bayesian}                            
\end{figure}

In NLOS channels, Fig. \ref{SERFourier_Bayesian} reports the impact of the limited number of DoF 
on A-ED. We observe that when the channel gains are equal, the performance is better than that of the case where the channel gains have an exponential decay power delay spectrum. The DoF of the channel is the limiting factor for the SER. The larger the DoF ($2L$), the better the performance  until it reaches the maximum DoF ($2M$) that is determined by the number of antennas. When $L = M$, we can see that the performance coincides with that of  Rayleigh fading channels. When $L \le M$, we observe an error floor: increasing the number of antennas beyond $L$ does not bring any benefits. We draw the conclusion that A-ED performs poorly when the number of paths $L$ is small.  

\begin{figure}
	\centering
	\resizebox{0.9\columnwidth}{!}{\includegraphics{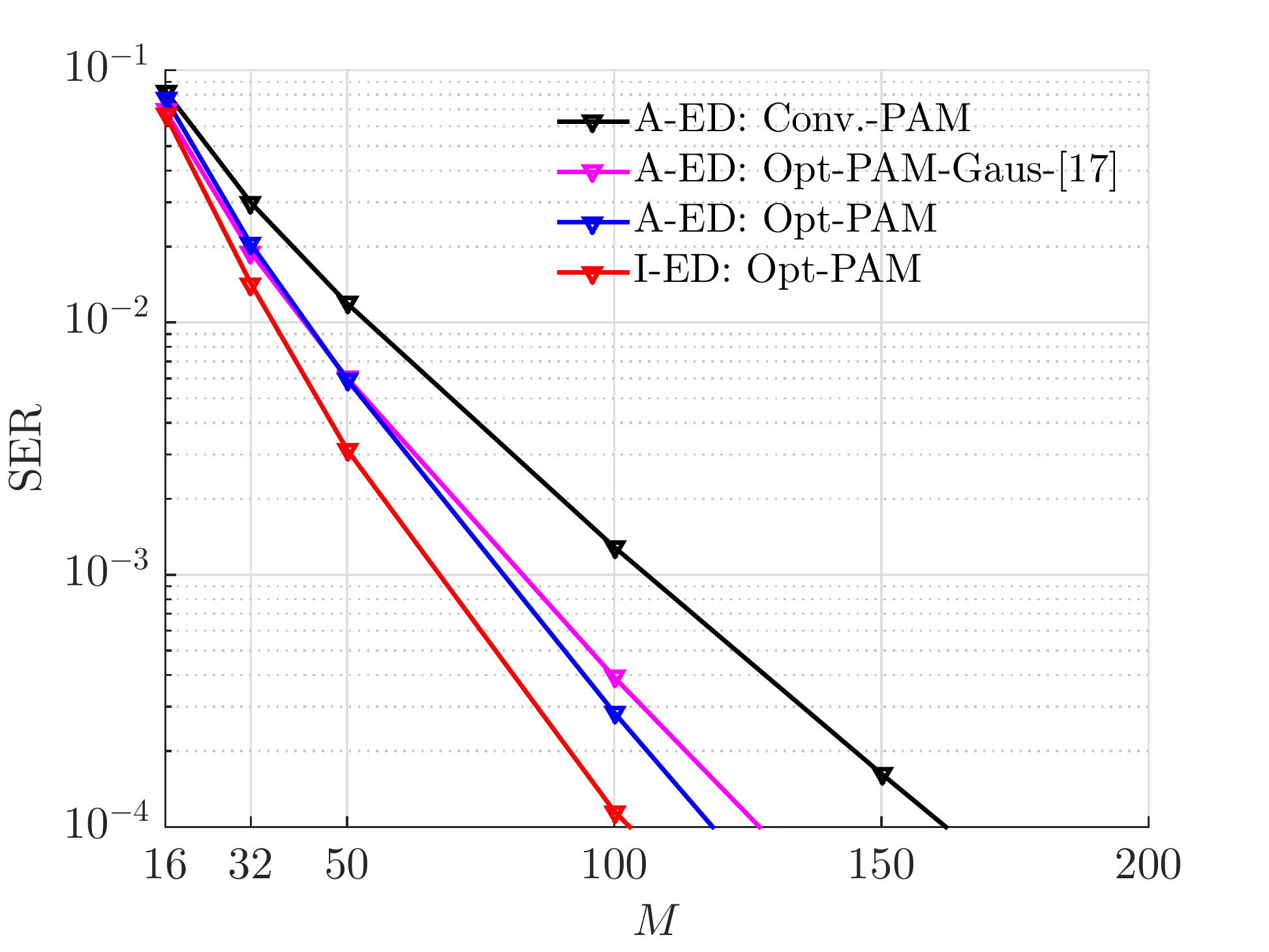}}
	\caption{Sparse LOS channels: SER comparison of optimized PAM and conventional (conv.) PAM with SNR $= \SI{4}{\dB}$, $L = 9$.}
	\label{SERFourier_BayesianLOS}                            
\end{figure}

\begin{figure}
	\centering
	\resizebox{0.9\columnwidth}{!}{\includegraphics{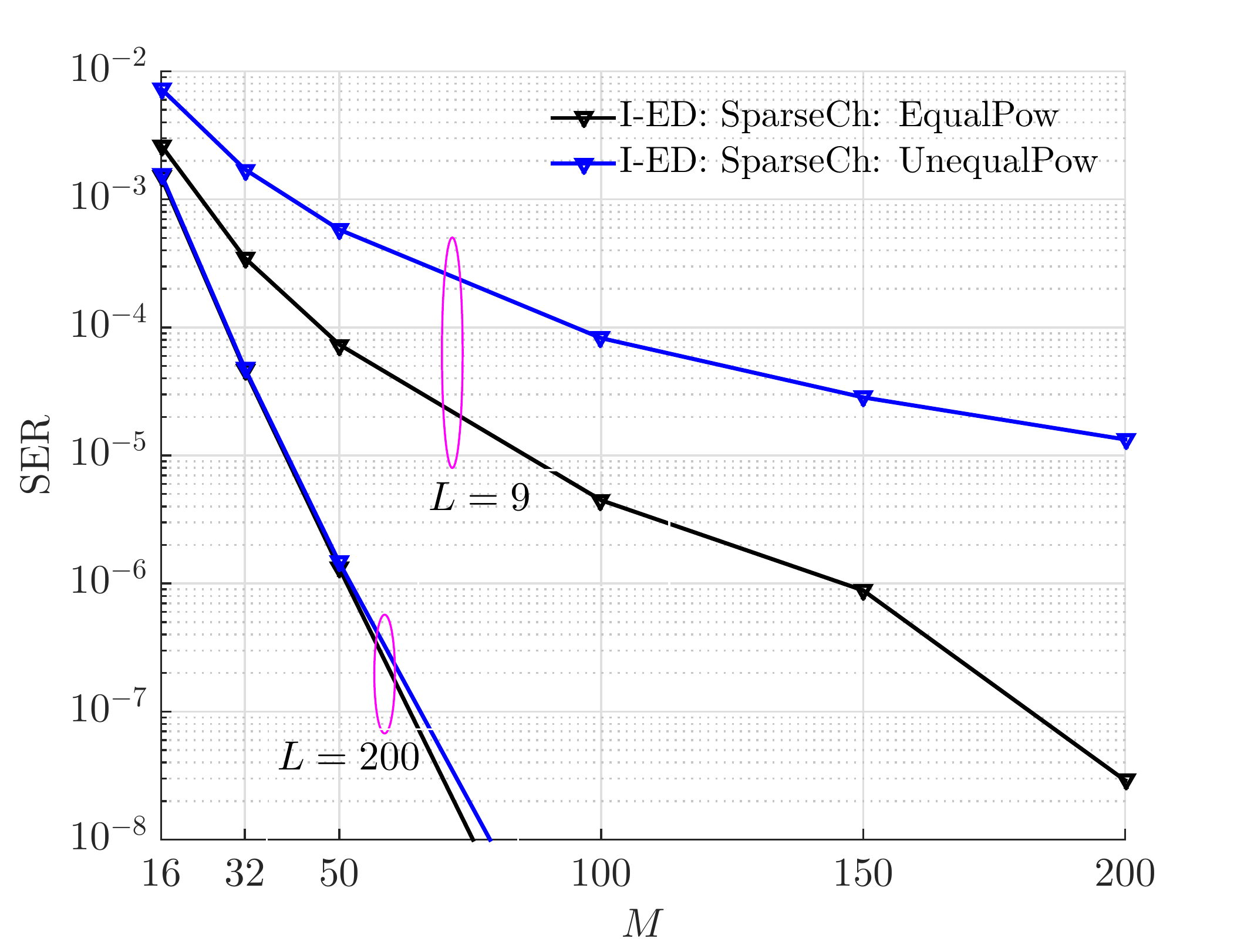}}
	\caption{SER versus the no. of antennas for different number of path components in NLOS sparse channels: Conventional PAM, SNR $= \SI{10}{\dB}$ and assuming instantaneous channel energy is known.}
	\label{SERFourier_Inst}                            
\end{figure}

As expected, I-ED significantly outperforms A-ED: see Fig. \ref{SERFourier_Inst}. When $L$ is small, the performance still improves when the number of receiver antenna increases. I-ED  leads to a SER around $1\times 10^{-4}$ at $M = 100$ and $L = 9$. Thus this method potentially can be employed in slow fading channels with a few number of path components. 


\section{Summary and Conclusions}
\label{sec:conc}

We propose two ED methods for a single stream transmission and reception by  a very large number of antennas. The first method I-ED is based on the instantaneous channel energy and exploits noise hardening. Based on the proposed decision thresholds, we show that the performance of I-ED asymptotically approaches the performance of coherent detection in high SNR regimes. The second method A-ED is based on the average channel energy and exploits both channel and noise hardening. For this method, an error floor is observed when PAM of order higher than two is employed due to channel energy uncertainties.  While A-ED gives satisfactory performance when the number of DoFs available in the channel is sufficiently high, its performance significantly degrades when the number of DoF becomes limited, such as in sparse channels. But the performance of I-ED is still promising in channels with large or limited number of DoFs, which can be of high relevance for practical applications. 



%

\section*{Acknowledgment}
The research presented in this paper is supported by the Danish Council for Independent Research (Det Frie Forskningsr\aa d) DFF $133500273$. 

\appendices
\section{Proof: Conventional PAM is Optimal at High SNR Regime when I-ED is Employed} \label{APPdix:OptimalConst}

We now prove that, at high SNR, conventional non-negative PAM is optimal in terms of minimizing the SER when I-ED is employed. Meanwhile, the optimal constellation also leads to an equalized $\gamma_{u,p}$ and $\gamma_{l,p}$ for $p = 0, \ldots, P-2$. 

When the SNR $\rho_h$ is sufficiently high, we have  $\gamma_{u,p} \approxeq \gamma_{l,p}$ in Section \ref{BF:analysis}. Thus,  computing the optimal constellation in \eqref{OptInst} is equivalent to solving
\begin{align} \label{OptInstEqui}
& \arg\min_{\sqrt{\epsilon_0}, \ldots, \sqrt{\epsilon_{P-1}}} \sum_{p = 0}^{P-2} Q (\sqrt{\gamma_{u,p}}) \\
&  \text{Subject to  } \frac{1}{P}\sum_{p = 0}^{P-1} \epsilon_p = 1, \notag \\      
& \hspace{50pt} \sqrt{\epsilon_i} < \sqrt{\epsilon_j} \; \text{when} \;\; i<j. \notag 
\end{align}

Invoking the Lagrange method and utilizing the fact that $$\frac{\partial Q (\gamma_{u,p})}{\partial \sqrt{\epsilon_p}} = -  \frac{1}{2 \pi} e^{-\frac{\gamma_{u,p}}{2}} \frac{\partial \sqrt{\gamma_{u,p}}}{\partial \sqrt{\epsilon_p}}, $$
we obtain that the conventional PAM is optimal.
Inserting the conventional PAM into \eqref{GammaRelations}, we readily obtain equalized $\gamma_{u,p}$ and $\gamma_{l,p}$ for $p = 0, \ldots, P-2$.

\section{OOK: Dominance of the SER Caused by $\epsilon_1$} \label{App1}
Since $
\delta_0^u =  \sqrt{\frac{  \rho'_1}{2}} $ and
$\delta_1^l = \frac{1}{\sqrt{2 \rho'_1 }}$,
plugging these two terms in~\eqref{PeuA} and~\eqref{PelA}, we obtain
\begin{equation}\label{PeuA1}
P_e^u(\sigma^{2}_h , \epsilon_0) \approx \left[ \sqrt{\frac{  \rho'_1}{2}} e^{\left(1- \sqrt{\frac{  \rho'_1}{2}}\right)} \right]^M
\end{equation}
\begin{eqnarray}\label{PelA1}
P_e^l (\sigma^{2}_h, \epsilon_1) \approx \left[ \frac{1}{\sqrt{2 \rho'_1 }} e^{\left(1- \frac{1}{\sqrt{2 \rho'_1 }}\right)} \right] ^M.
\end{eqnarray}
When $\rho'_1 \gg 1$, $\sqrt{\frac{  \rho'_1}{2}} e^{\left(1- \sqrt{\frac{  \rho'_1}{2}}\right)} \approx \sqrt{\frac{  \rho'_1}{2}} e^{- \sqrt{\frac{  \rho'_1}{2}}} $ and $\frac{1}{\sqrt{2 \rho'_1 }} e^{\left(1- \frac{1}{\sqrt{2 \rho'_1 }}\right)} \approx \frac{1}{\sqrt{2 \rho'_1 }} e$. As a result,  the  error probability for $\epsilon_0$ and $\epsilon_1$  diminishes as $\rho \rightarrow + \infty$. Meanwhile, the SER for $\epsilon_0$ decays at a much faster scale due to the fact that the ratio between the inner arguments of \eqref{PeuA1} and \eqref{PelA1} are much smaller than $1$: $$\frac{\sqrt{\frac{  \rho'_1}{2}} e^{- \sqrt{\frac{  \rho'_1}{2}}}}{\frac{1}{\sqrt{2 \rho'_1 }} e} \approx \rho'_1 e^{-\sqrt{\frac{  \rho'_1}{2}}}$$ when $\rho $ is sufficiently large.


\bibliographystyle{IEEEtran}
\bibliography{IEEEabrv,AAURef}

\end{document}